\documentstyle[twocolumn,aps,epsfig]{revtex}
\input{epsf.tex}
\begin{document}
\title{Solitons and vortices in ultracold fermionic gases}
\author{Tomasz Karpiuk,$^\dag$ Miros{\l}aw Brewczyk,$^\dag$ and
        Kazimierz Rz\c a\.zewski$^\ddag$}
\address{$^\dag$ Uniwersytet w Bia{\l}ymstoku, ul. Lipowa 41, 
                 15-424 Bia{\l}ystok, Poland}
\address{$^\ddag$ Centrum Fizyki Teoretycznej PAN and 
                  College of Science,\\
                  Al. Lotnik\'ow 32/46, 02-668 Warsaw, Poland}

\date{\today}

\maketitle

\begin{abstract}

We investigate the possibilities of generation of solitons and
vortices in a degenerate gas of neutral fermionic atoms.
In analogy with, already experimentally demonstrated, technique 
applied to gaseous Bose-Einstein condensate we propose the phase
engineering of a Fermi gas as a practical route to excited states
with solitons and vortices. We stress that solitons and vortices
appear even in a noninteracting fermionic gas. For solitons, in a 
system with sufficiently large number of fermions and appropriate
trap configuration, the Pauli blocking acts as the interaction between 
particles. \\

PACS number(s): 05.30.Fk, 03.75.Fi  \\

\end{abstract}

Recent experimental achievement of quantum degeneracy in a dilute
gas of fermionic $^{40}$K atoms \cite{Jin} has triggered theoretical
interest in properties of such systems. In particular, the
interactions between ultracold fermions were studied both in the
static and dynamic context. Since Fermi atoms in the same spin state
do not interact via s--wave collisions, having atoms in two
hyperfine states is necessary to perform cooling of a system using
evaporative technique. That is the idea implemented in experiment
\cite{Jin}. In Ref. \cite{Bruun} in phase collective excitations
of such two--component system are discussed within the hydrodynamic
limit and method for detection of the onset of Fermi degeneracy is 
proposed. Static properties of two--component system were already
analyzed in Ref. \cite{Bruun1}. In a one--component gas in the
absence of s--wave collisions other types of forces are getting
important, for example, the dipole--dipole forces. The properties
of trapped fermionic dipoles, including the stability analysis,
have been investigated recently in Ref. \cite{Krzysiek}. Interesting 
results on the transition temperature to BCS phase and the detection 
of Cooper pairing were also obtained \cite{BCS}.

In this Letter, we investigate another aspect of dynamic behavior
of a degenerate Fermi gas. Following recent success in experimental
demonstration of topological defects in Bose--Einstein condensate
such as vortices \cite{JILA,Dalibard} we ask the question whether
vortices can be observed in a gas of neutral fermionic atoms. We are 
interested in a generation of solitons and vortices in a normal state 
of a Fermi gas, i.e., above the temperature for a BCS--type phase 
transition. We find that the phase imprinting technique \cite{Maciek,NIST} 
already applied for the Bose--Einstein condensate also works for an ultracold 
Fermi gas, although the nature of vortices appears to be different. 
In fact, because of technical reason only dark solitons were observed in the 
Bose--Einstein condensate so far by using this method. Then, we analyze 
the ``birth'' and dynamics of solitons in a fermionic gas too and discover 
some differences in comparison with bosons.

As opposed to the Bose--Einstein condensate, the fermionic system
possesses no macroscopic wave function. This fact does not exclude,
however, the existence of excited states of such systems with
solitons and vortices. To proof that, we have solved the many--body 
Schr\"odinger equation for a one--dimensional and three--dimensional 
noninteracting Fermi atoms in a harmonic trap. At zero temperature
the many--body wave function is given by the Slater determinant with 
the lowest available one--particle orbitals occupied. By using the 
sufficiently fast phase imprinting technique each atom acquires the same 
phase $\phi(\vec{r})$, hence the orthogonality of one--particle orbitals 
is not broken. Time (unitary) evolution does not spoil this orthogonality too, 
so the diagonal part of one--particle density matrix is always the sum of 
one--particle orbitals densities during the evolution.

\begin{figure}[t!]    
\centerline{\epsfig{file=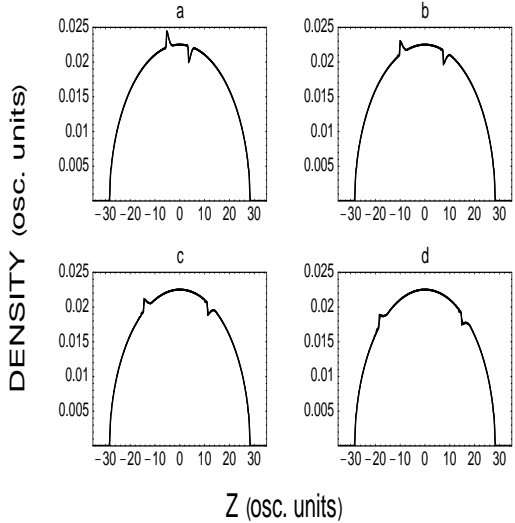,height=2.8in,width=2.8in}}
\vspace{10pt}
\caption{Density profiles of 1D noninteracting fermionic gas at
different times: (a) 0.16, (b) 0.31, (c) 0.47, and (d) 0.63 in 
units of $1/\omega$  after imprinting a single phase step of
$\phi_0=2.0\,\pi$ and $\zeta=0.5$ \mbox{osc. units}. Note that each graph 
consists actually of two curves: one obtained based on the many--body 
Schr\"odinger equation (solid line) and the second coming from the 
one--dimensional Thomas--Fermi model (dashed line).}
\label{1Dsol}
\end{figure}

In one--dimensional case only generation of solitons can be considered.
We take the phase being imprinted in the form
$\phi(z)=\phi_0 (1+\tanh(z/\zeta))/2$ with the phase jump determined 
by $\phi_0$ and the width of the phase jump equal to $\zeta$. In Fig.
\ref{1Dsol} we plotted the density profiles of a system of N=$400$
noninteracting Fermi atoms at various times. It is clear that two 
solitons, the dark and the bright one, are generated. They propagate
in opposite directions with the speed of bright soliton being higher. 
After reaching the edge of the system the solitons return to the center
of a trap and oscillatory motion of solitons is observed (oscillatory
motion of solitons in Bose--Einstein condensate was reported in Ref.
\cite{osc_sol}). It is important to note that in each graph of Fig.
\ref{1Dsol} we show two curves. One of them is a result of a solution 
of many--body Schr\"odinger equation. The second comes from a 
one--dimensional Thomas--Fermi approximation (discussed below). The perfect 
agreement between an exact and approximate solutions certainly proves the 
validity of the Thomas--Fermi approach in one dimension for sufficiently 
large number of atoms. It also proves the existence of solitons in the 
system of noninteracting fermions suggested recently in Ref. \cite{1DFermi}. 
It means also that for N=$400$ atoms and more, the Pauli blocking plays 
the role of the interaction between atomic fermions and assures the local 
equilibrium necessary within the Thomas--Fermi approximation. 

Since the harmonic potential is a special one in a sense that it allows 
for a propagation of wave packets without spreading, we have also 
checked the case of a system of fermions confined in a one--dimensional 
box. Again, after phase imprinting we found two well--defined pulses 
propagating in opposite directions with distinct velocities. The speed 
of the bright soliton is higher than the speed of sound in a fermionic 
gas at T=0 temperature (given by $(\hbar \pi/m) \,\rho$) whereas the dark
soliton propagates slower than the sound wave.
It is important to verify the above scenario in more than
one--dimensional space. To this end, let's consider a noninteracting 
Fermi gas in an arbitrary three--dimensional harmonic trap 
characterized by the frequencies $\omega_x$, $\omega_y$, and $\omega_z$. 
One--particle orbitals are taken as a product $\, \varphi_{n_x}^{(1)} (x) 
\, \varphi_{n_y}^{(2)} (y) \, \varphi_{n_z}^{(3)} (z)$ of eigenvectors 
of the Hamiltonians of one--dimensional harmonic oscillators. The phase 
is being imprinted in the same form as in one--dimensional case. Since 
the Hamiltonian separates in coordinates 'x', 'y', and 'z', the time 
propagation of each orbital after phase imprinting can be easily reduced 
to one--dimensional problem:
\begin{eqnarray*}
\Psi_{n_x n_y n_z}(x,y,z,t) &=& e^{- i \rm{E}_{n_x} t /\hbar} \:
                          e^{- i \rm{E}_{n_y} t /\hbar}   \\
&\times& \varphi_{n_x}^{(1)} (x,0) \, \varphi_{n_y}^{(2)} (y,0) \\
&\times& \varphi_{n_z}^{(3)} (z,t)
\end{eqnarray*}
where
\begin{eqnarray*}
\varphi_{n_z}^{(3)} (z,t) &=& e^{- i \rm{H_z} \,t /\hbar}
\varphi_{n_z}^{(3)} (z,0) \, e^{i \phi(z)}
\end{eqnarray*}
\noindent
and $\rm{H_z}$ is the Hamiltonian of one--dimensional harmonic oscillator. 
Now the diagonal part of one--particle density matrix is given by the 
expression:
\begin{eqnarray}
\rho(\vec{r}, t) =
\frac{1}{\rm{N}} \, \sum_{ \rm{E}_{n_x n_y n_z} \leq \rm{E}_F}  \,
|\varphi_{n_x}^{(1)} (x,0)| ^2 \;
|\varphi_{n_y}^{(2)} (y,0)| ^2  \nonumber \\ \nonumber \\
\times \, |\varphi_{n_z}^{(3)} (z,t)| ^2  \phantom{aaaaaaaaaa}\, & &
\label{3Dden}
\end{eqnarray}
where $\rm{N}$ is the number of atoms and sum is performed over the
one--particle states below Fermi energy.

\begin{figure}[b!]    
\centerline{\epsfig{file=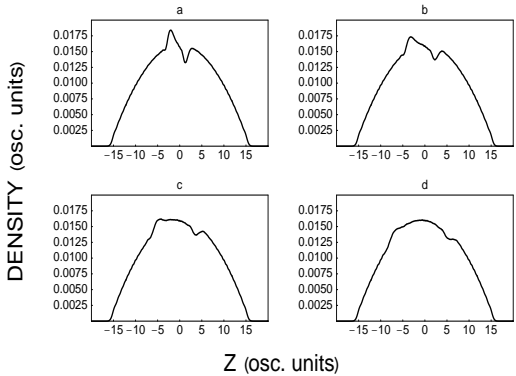,height=2.0in,width=2.8in}}
\vspace{10pt}
\caption{Evolution of the density distribution of $1000$ Fermi atoms
in three--dimensional harmonic trap ( $\omega_x=2\,\omega_z$ and 
$\omega_y=64\,\omega_z$) after imprinting a single phase step of 
$\phi_0=1.0\,\pi$ and $\zeta=0.2$ \mbox{osc. units}. The successive 
frames correspond to moments: (a) 0.25, (b) 0.40, (c) 0.60, and 
(d) 0.85 in units of $1/\omega_x$.}
\label{3Dcut}
\end{figure}

\begin{figure}[b!]    
\centerline{\epsfig{file=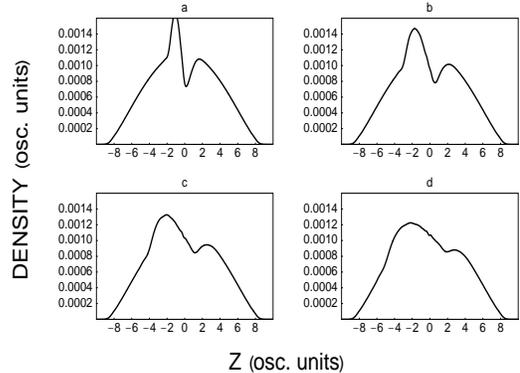,height=2.0in,width=2.8in}}
\vspace{10pt}
\caption{Evolution of the density distribution of $10^4$ Fermi atoms
in three--dimensional spherically symmetric harmonic trap after 
writing a single phase step of $2.0\,\pi$ and $\zeta=0.5$ 
\mbox{osc. units}. The successive frames correspond to moments: 
(a) 0.1, (b) 0.2, (c) 0.3, and (d) 0.4 in units of $1/\omega$.}
\label{3Dsscut}
\end{figure}

One of the obvious regimes where the Thomas--Fermi approximation holds 
is that of a cigar--shaped trap elongated enough, depending on the number 
of confined atoms. For example, in the case of a ground state of N=$400$ 
fermions in the harmonic potential with oscillator frequencies taken as 
$\omega_x=\omega_y=400\omega_z$ only the smallest possible radial quantum 
numbers are involved: $n_x=n_y=0$. It follows from the formula (\ref{3Dden}) 
that the density along 'z' axis (or any line parallel to it) is just 
one--dimensional density multiplied by a constant. Certainly, the overall 
behavior found in one--dimensional case translates to three-dimensional
one. Conditions presented above could be considered also as a criterion 
for treating the real Fermi gas as a one--dimensional system.
\begin{figure}[b!]    
\centerline{\epsfig{file=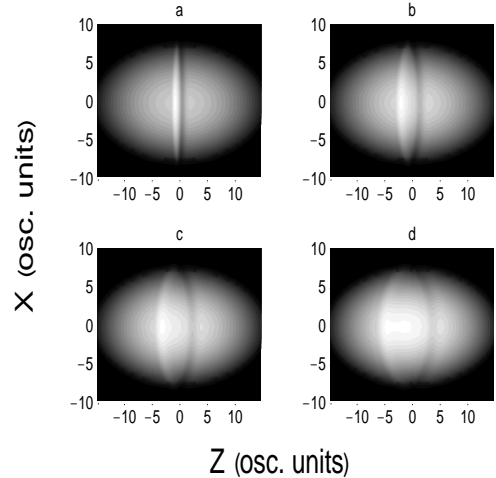,height=2.8in,width=2.8in}}
\vspace{10pt}
\caption{Images of the fermionic density of N=$1000$ atoms after 
writing a single phase step of $1.0\,\pi$ and $\zeta=0.2$ 
\mbox{osc. units} onto a three--dimensional noninteracting Fermi gas 
confined in a trap configured as follows: $\omega_x=2\,\omega_z$ and 
$\omega_y=64\,\omega_z$. The snapshots are taken at times of (a) 0.10, 
(b) 0.25, (c) 0.40, and (d) 0.60 in units of $1/\omega_x$ .}
\label{3Dsol}
\end{figure}
Now we start to depart from ``one--dimensional'' geometry.
In Figs. \ref{3Dsol} and \ref{3Dcut} we show the contour plots 
of density in the 'xz' plane and the density profiles along 'z' 
direction respectively after writing a single phase step onto a 
three--dimensional gas, although confined in a disk--shaped trap.
The system behaves effectively as a two--dimensional gas; the number
of atoms and the fundamental frequencies are chosen in such a way
that one ($n_y$) of the quantum numbers equals zero. In this case
we again observe two pulses moving in opposite directions, however
they are broader in comparison with one--dimensional case (see Figs.
\ref{1Dsol} and \ref{3Dcut}). At the same time results obtained based
on the Thomas--Fermi approximation differ qualitatively, showing the
existence of the second dark soliton. Only going to much larger number
of atoms improves the agreement between both approaches, however then
all structures are getting less pronounced. In three-dimensional
spherically symmetric trap we have never observed double pulse
behavior. Instead of that, the dark soliton and the density wave are 
generated (see Fig. \ref{3Dsscut}) much as in the case of solitons
generated in the Bose--Einstein condensate \cite{Maciek}. The dark 
soliton propagates with the velocity smaller than the speed of sound 
and experiences (as in one--dimensional case) the oscillatory motion.

A convenient way to discuss Thomas--Fermi approximation is to 
start from a set of equations for reduced density matrices. Since there 
is no interaction between particles, the equation of motion for the 
one--particle density matrix does not involve the two--particle density 
matrix $\rho_2$ and is given by
\cite{equ} %
\begin{eqnarray}
i\hbar \frac{\partial}{\partial t} \rho_1(\vec{r}_1,\vec{r}_2,t) = 
-\frac{\hbar^2}{2 m} (\vec{\nabla}_1^2 - \vec{\nabla}_2^2) \,
\rho_1(\vec{r}_1,\vec{r}_2,t) \phantom{11111} & &
\nonumber \\
+ \left[V_{ext}(\vec{r}_1,t)-V_{ext}(\vec{r}_2,t)\right]
\rho_1(\vec{r}_1,\vec{r}_2,t) \;, \phantom{11111111111l}  & &
\label{density}
\end{eqnarray}
where $V_{ext}(\vec{r},t)$ is the external potential. In the
limit $\vec{r}_1 \rightarrow \vec{r}_2$ Eq. (\ref{density}) leads
to the continuity equation
\begin{eqnarray}
\frac{\partial \rho(\vec{r},t)}{\partial t} + \vec{\nabla} \cdot
\left[\rho(\vec{r},t) \, \vec{v}(\vec{r},t)\right] & = &  0  \;,
\label{hydr1}
\end{eqnarray}
where the density and velocity fields are defined as follows:
\begin{eqnarray}
\rho(\vec{r},t) & = & \lim_{\vec{r}_1 \rightarrow \vec{r}_2}
\rho_1(\vec{r}_1,\vec{r}_2,t) \nonumber \\
\vec{v}(\vec{r},t) & = & \frac{\hbar}{2 m}
\lim_{\vec{r}_1 \rightarrow \vec{r}_2}
(\vec{\nabla}_1 - \vec{\nabla}_2) \,
\chi_1(\vec{r}_1,\vec{r}_2,t)
\label{def}
\end{eqnarray}
and $\chi_1(\vec{r}_1,\vec{r}_2,t)$ is the phase of the one--particle
density matrix.

One can also rewrite Eq. (\ref{density}) introducing the 
center--of--mass ($\vec{R}=(\vec{r}_1+\vec{r}_2)/2$) and the 
relative position ($\vec{s}=\vec{r}_1-\vec{r}_2$) coordinates.
By taking the derivative of  Eq. (\ref{density}) with respect 
to the coordinate $\vec{s}$ the hydrodynamic Euler--type equation
of motion is obtained in the limit $\vec{s} \rightarrow 0$
\begin{eqnarray}
\frac{\partial \vec{v}(\vec{r},t)}{\partial t} & = & 
-\frac{\vec{\nabla} \cdot T}{m\, \rho(\vec{r},t)}
-\frac{1}{2} \vec{\nabla} \vec{v}^{\, 2}(\vec{r},t)
-\frac{\vec{\nabla} V_{ext}(\vec{r},t)}{m}  \;.
\label{hydr2}
\end{eqnarray}
Here, the kinetic--energy stress tensor $T$, whose elements are
given by
\begin{eqnarray*}
T_{kl} & = & -\frac{\hbar^2}{m}
\lim_{\vec{s} \rightarrow 0}
\frac{\partial^{\,2} \sigma_1(\vec{R},\vec{s},t)}
{\partial s_k \partial s_l}  \;,
\end{eqnarray*}
depends on the one--particle density matrix (more precisely on its 
amplitude $\sigma_1(\vec{r}_1,\vec{r}_2,t)$) and is calculated 
based on a local equilibrium assumption which is the substance
of Thomas--Fermi approximation.

We have solved the set of Eqs. (\ref{hydr1}) and (\ref{hydr2}) with 
kinetic--energy stress tensor approximated by its Thomas--Fermi 
form, i.e., diagonal tensor with (in three--dimensional case)
$(\hbar^2/m) (1/30/\pi^2) (6 \pi^2 \rho)^{5/3}$ term at each position.
In fact, we applied the inverse Madelung transformation \cite{Madelung}
to Eqs. (\ref{hydr1}) and (\ref{hydr2}) and used the Split-Operator
method combined with the imaginary--time propagation technique 
(for generation of the ground state of the trapped Fermi gas) as well
as the real--time propagation (for phase imprinting and analysis of
dynamic instabilities after that).

%
%
By using the definition (\ref{def}) of the velocity field one can easily 
calculate
\begin{eqnarray*}
\vec{v}(\vec{r},t) & = & \frac{\sum_{i=1}^{\rm{N}} \, \rho_i(\vec{r} ,t)
\, \vec{v}_i (\vec{r},t)} {\sum_{i=1}^{\rm{N}} \, \rho_i(\vec{r} ,t)} \, .
\end{eqnarray*}
Although the one--particle velocities $\vec{v}_i$ are irrotational,
the global velocity is, in principle, rotational. Therefore, for an
ultracold Fermi gas in a normal phase one should not expect the
existence of quantized vortices. However, as we'll see, the excitations
of fermionic gas with a hole in a density along some line (or a dot in
two--dimensional case) and non--zero circulation around this line are 
still possible.

\begin{figure}[t!]    
\centerline{\epsfig{file=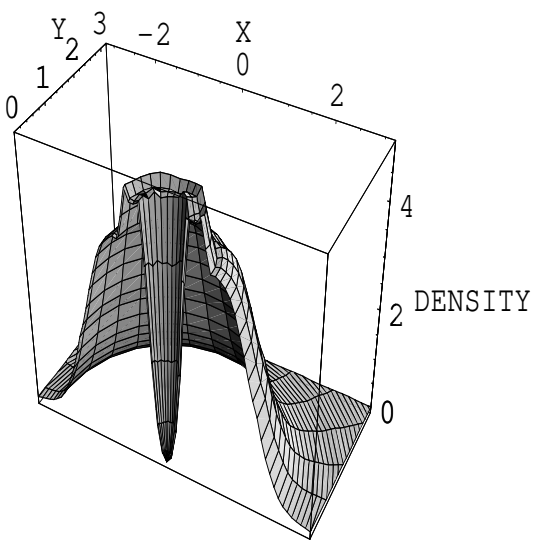,height=2.8in,width=2.8in}}
\vspace{10pt}
\caption{Density (in arbitrary units) of the system of N=$6$ 
fermions in the excited state with a vortex.}
\label{2Dvor}
\end{figure}

To discuss the possibility of generation of vortices in a degenerate
Fermi gas we involve the polar coordinates. Hence, the eigenvectors 
are given by:
$\Psi_{n m} (r,\phi) = A \, r^{|m|} \: L_n^{|m|} (r^2) \:
                 e^{r^2 /2} \, e^{i m \phi}$     and
$A^2 = n! / [\pi (n+|m|)!]$.
As in Refs. \cite{Dobrek,IFT}, we pass a laser pulse through the
appropriately tailored absorption plate before impinging it on the
atomic gas. The laser pulse is short with a duration of the order
of the fraction of microsecond and the frequency is detuned far
from the atomic transition (only the dipole force is important).
Under such conditions the atomic motion is negligible during the
pulse and the only effect of light on the atoms is imprinting the 
phase. To generate a vortex one has to prepare the absorption plate 
with azimuthally varying absorption coefficient. Changing the laser 
intensity and the duration of a pulse (the laser pulse area) one 
is able to design various vortex excitations of the Fermi gas. 
The state right after phase imprinting turns out to be dynamically 
unstable and soon after a vortex at the center of the trap is formed.
In Fig. \ref{2Dvor} we show the density after the pulse with the
laser pulse area equal to $3\times 2\pi$ was applied to the system of N=$6$
fermions. The stable, not quantized vortex is generated. Increasing the
number of atoms leads to the lower contrast of the vortex but 
increasing (keeping the same number of atoms) the pulse area
restores the contrast.

In conclusion, we have shown that the phase imprinting method can 
be employed for generation of solitons and vortices in ultracold 
fermionic gases. We have demonstrated that idea explicitly by solving 
the many--body Schr\"odinger equation for a gas of noninteracting 
neutral fermionic atoms. There exists differences in comparison with 
the Bose--Einstein condensate case. For example, the presence of bright 
solitons moving with the velocity higher than the speed of sound or the 
nature of the vortices which are not quantized. All these phenomena present 
for fermionic gases have nothing to do with the potential formation of 
bound states (Cooper pairs) followed by the condensation, since no 
interaction between Fermi atoms was included in the calculations.

\acknowledgements
K.R. acknowledges support of the Subsidy by Foundation for 
Polish Science.

\end{document}